\documentclass[]{hdsr}

\usepackage{tabulary}
\usepackage{tabularx}
\usepackage{url}

\graphicspath{{figs/}}

\begin{document}

\newgeometry{bottom=1.5in}

\begin{center}

  \title{The evolving ecosystem of COVID-19 contact tracing applications}
  \maketitle

  \thispagestyle{empty}
  
  \vspace*{.2in}

  \begin{tabular}{cc}
    Benjamin Levy\upstairs{\affilone,*}, Matthew Stewart\upstairs{\affilone,*}
   \\[0.25ex]
   {\small \upstairs{\affilone} Harvard University, John A. Paulson School of Engineering and Applied Sciences} \\
  \end{tabular}
  
  \emails{\upstairs{*}Corresponding authors. Both authors contributed equally.\newline benjaminlevy@g.harvard.edu, matthew\_stewart@g.harvard.edu
    }
  \vspace*{0.4in}

\hypertarget{Abstract}{%
\section*{Abstract}\label{Abstract}}

Since the outbreak of the novel coronavirus, COVID-19, there has been increased interest in the use of digital contact tracing as a means of stopping chains of viral transmission, provoking alarm from privacy advocates. Concerning the ethics of this technology, recent studies have predominantly focused on (1) the formation of guidelines for ethical contact tracing, (2) the analysis of specific implementations, or (3) the review of a select number of contact tracing applications and their relevant privacy or ethical implications. In this study, we provide a comprehensive survey of the evolving ecosystem of COVID-19 tracing applications, examining 152 contact tracing applications and assessing the extent to which they comply with existing guidelines for ethical contact tracing. The assessed criteria cover areas including data collection and storage, transparency and consent, and whether the implementation is open source. We find that although many apps released early in the pandemic fell short of best practices, apps released more recently, following the publication of the Apple/Google exposure notification protocol, have tended to be more closely aligned with ethical contact tracing principles. This dataset will be publicly available and may be updated as the pandemic continues.
\end{center}

\vspace*{0.15in}
\hspace{10pt}
  \small	
  \textbf{\textit{Keywords: }} {COVID-19, contact tracing, exposure notification, ethics, privacy}
  
\copyrightnotice

\section*{Media Summary}

Contact tracing is an essential epidemiological tool in the fight against COVID-19. However, the magnitude and transmissibility of the disease has far outstripped the capacity of health authorities to track, trace, and treat infections as they arise. In response, digital contact tracing applications have emerged as a means to augment traditional manual contact tracing. Beginning with Singapore’s TraceTogether — a Bluetooth-based app that gained significant attention as a part of the government’s pandemic strategy — the number of digital contact tracing applications has exploded over the course of the pandemic. Concurrently, privacy advocates have been raising the alarm that these apps may lead to an erosion of individual liberties. 

In this article, we provide a comprehensive review of digital contact tracing applications around the world. We collect data on 152 applications from 77 different countries and evaluate those applications according to their compliance with established ethical principles outlined by the World Health Organization. We find that there is considerable heterogeneity between regions, countries, and apps in terms of their adherence to best practices. Notably, the release of Apple and Google’s exposure notification protocol seems to have encouraged a movement towards applications with stronger privacy protections. Although this is an encouraging development, it is difficult to determine whether these applications have had or will have an impact on the spread of the disease. Therefore, we provide an open dataset of these applications, with the hope that subsequent work may be better situated to evaluate their efficacy and uptake to help inform the implementation of future digital public health interventions.

\hypertarget{introduction}{%
\section*{Introduction}\label{introduction}}

The outbreak of SARS-CoV-2, the virus responsible for COVID-19, has resulted in an ongoing global pandemic that has, as of February 1, 2021, infected over 100 million people and
killed more than 2 million people \citep{JohnHopkins}. As an early response, many countries implemented temporary stay-at-home measures compelling non-essential workers to remain indoors under quarantine, as well as enforcing social distancing measures and the wearing of masks in public places. While these measures proved effective in the short run when met with high public adherence levels, they have not proved a feasible and effective long-term solution for controlling the virus's spread in most countries. Full lockdown measures come with significant economic, social, educational, psychological, and medical ramifications; for instance, unemployment in May in the United States was estimated to be as high as 16\% as a result of the pandemic \citep{kochhar2020unemployment}. A practical and proven method for counteracting the spread of transmissible diseases is through contact tracing: using social networks to determine contacts who have had recent interactions with an infected individual. These contacts can then be quarantined, tested, or instructed to self-isolate for 14 days, the putative upper bound on the incubation period of the virus \citep{linton2020incubation}.

Traditional contact tracing relies on many human workers to track down and reach out to each potential contact. It can therefore be prohibitively expensive, slow, and challenging to scale \citep{hart2020outpacing}. Many countries and private organizations turned to smartphones as an alternative source for personal contact history. There has been a rapid development of mobile applications designed to aid contact tracing to suppress transmission of COVID-19. These applications typically use GPS or Bluetooth data to detect a device's proximity to nearby devices. Although manual contact tracing is expected to have higher effectiveness at reducing transmission rates than digital contact tracing for a fixed number of individuals \citep{lancet2020}, the scalability afforded by digital contact tracing coupled with its relatively low cost makes large-scale implementations more feasible than manual contact tracing.

Naturally, the prospect of large-scale monitoring of the public using contact tracing applications has alarmed privacy advocates \citep{aclu2020whitepaper}. The hasty implementations of such systems in response to the pandemic's rapid proliferation and severity raised concerns about whether potential ethical and privacy implications were sufficiently considered during the rollout of applications. Further privacy concerns have arisen from contact tracing applications utilizing geolocation data, centralized storage systems, closed source implementations, and government coercion \citep{shukla2020privacy}. Consumer advocates fear that a disproportionate emphasis on health over privacy could undermine the protection of civil liberties far beyond the stated intentions of these applications, similar to the vast expansion of state surveillance in the United States post-9/11 \citep{hart2020outpacing, guariglia2020eff}. In turn, such concerns may negatively impact adoption rates and thus reduce the effectiveness of contact tracing. Simulations have shown that the efficacy of contact tracing applications is highly dependent on adoption rates, with a level around 50-60\% necessary to limit the spread of COVID-19 \citep{ferretti2020quantifying}. Survey results from Canada and the United States have supported the notion that privacy is a significant reason people may be hesitant to use digital contact tracing apps \citep{rheault_musulan_2020, zhang2020americans}, and has remained consistent throughout the pandemic \citep{simko2020covid19}. Although polls have demonstrated a high willingness to download contact tracing apps \citep{Altmann2020.05.05.20091587}, short of coercion, maximizing adoption rates can only be achieved if individuals do not perceive the application to encroach unreasonably upon civil liberties. While there may seemingly be exceptions to this rule, such as South Korea's government-backed centralized approach, deviations may be at least partially explained by factors unique to South Korea, such as (1) societal support for government interventions due to prior exposure to outbreaks like Middle East Respiratory Syndrome (MERS), and (2) rapid implementation that had little time to garner negative press and may have instead dampened uptake in other countries \citep{park2020information}.

Digital contact tracing is inherently multidisciplinary and draws upon public health, epidemiology, computer science, and data science. Each of these fields contains multiple sets of ethical guidelines. Several attempts have been made to combine these to form an appropriate set of guidelines for ethical contact tracing \citep{morley2020ethical, Parker427, who2020ethical}. However, the difficulty in realizing these guidelines in practice lies in the fact that there are already many existing contact tracing applications (at least 124 released, by our estimation) and a multitude of public, private, and academic actors involved in their development. These factors can lead to a lack of coordination due to these various actors' competing interests and discordant practices, resulting in no universally observed standard for ethical contact tracing. Although vaccines are now available and being distributed to the public, there is still a lack of consensus and unanswered questions about how future contact tracing programs should be conducted to balance the utilization of data and preservation of individual privacy. In this paper, we conduct an exhaustive survey of contact tracing applications and assess the adherence of these applications to established ethical contact tracing guidelines. Our main contributions are a database of contact tracing applications, an assessment of how well these applications conform with the aforementioned ethical principles based on their synthesis into a single 8-point scale (which we denote the ``Ethical Alignment Score''), and a comparative analysis of applications based on location, release date, and other variables of interest.

\restoregeometry
\newgeometry{bottom=0.5in}

\hypertarget{methods}{%
\section*{Data collection and coding}\label{methods}}

In their interim guidance report on ethical contact tracing, the World Health Organization (WHO) outlined 17 principles to which contact tracing applications should adhere \citep{who2020ethical}. These guidelines dovetail with other ethical contact tracing guidelines \citep{morley2020ethical, Parker427} and provide a set of criteria upon which existing contact tracing applications can be assessed. The proposed principles are summarized under the following 17 categories: time limitation, testing and evaluation, proportionality, data minimization, use restriction, voluntariness, transparency and explanability, privacy-preserving data storage, security, limited retention, infection reporting, notification, tracking of COVID-19-positive cases, accuracy, accountability, civil society and public engagement, and independent oversight. For a more detailed explanation of these categories, we refer the reader to the original WHO article (\citeyear{who2020ethical}). The remainder of this paper uses these guidelines as a basis for a set of questions that can be answered with publicly available data about each contact tracing application. The responses to these questions are combined into a single value, which we denote the Ethical Alignment Score.

We compiled a list of COVID-19 contact tracing applications by conducting independent research on applications found in following sources:

\begin{enumerate}
\item
  \href{https://en.wikipedia.org/wiki/COVID-19_apps}{Wikipedia page on
  COVID-19 apps}
\item
  \href{https://privacyinternational.org/examples/apps-and-covid-19}{Privacy International}
\item
  \href{https://docs.google.com/document/d/16Kh4_Q_tmyRh0-v452wiul9oQAiTRj8AdZ5vcOJum9Y/edit?ts=5e801c37\#}{Open
  source community document on COVID-19 contact tracing technologies
  (Stop-COVID.tech, COVIDWatch, Mitra Ardron)}
  \item \href{https://www.privacy.org.nz/assets/2020-05-12-OPC-Comparison-of-COVID-19-Apps-colours.pdf}{New Zealand Office of the Privacy Commissioner: Overview of COVID-19 Contact Tracing Apps – 12 May 2020}
  \item \href{https://github.com/amnestytech/covid19-apps}{Amnesty International}
  \item \href{https://github.com/shankari/covid-19-tracing-projects}{Crowdsourced list of projects related to COVID-19 contact tracing}
  \item \href{https://flo.uri.sh/visualisation/2241702/embed}{MIT Technology Review Covid Tracing Tracker}
  \item \href{https://www.xda-developers.com/google-apple-covid-19-contact-tracing-exposure-notifications-api-app-list-countries/}{list of countries using Google and Apple's Contact Tracing API (XDA Developers)}
  \item \href{https://developers.google.com/android/exposure-notifications/apps}{Publicly-available Exposure Notifications apps (Google API)}
  \item \href{https://fipra.com/europe-covid-19-tracing-app-tracker/?intro_read=yes}{FIPRA Europe COVID-19 Tracing App Tracker}
  \item Google searching for terms ``covid-19'', ``contact tracing'', ``bluetooth contact tracing app'', ``gps contact tracing app''
\end{enumerate}

Using these resources and the materials put out by the makers of the applications themselves, we extracted a set of features for each app (when information was available) based on a subset of the ethical principles set out by the ethical contact tracing guidelines outlined previously (see table \ref{table:ethical-alignment-score}). We collected application release dates from \url{appannie.com}, a service for distributing statistics on mobile applications and websites, and approximate download numbers from the Google Play Store, when available.

\begin{table}[h]
\caption{Principles and definitions for scoring apps. These principles are based on Morley et al. \citep{morley2020ethical} and principles from the \citet{who2020ethical}.}
\vspace*{0.2in}
\begin{tabulary}{0.85\textwidth}{p{3cm}LL}
    \toprule
    \label{table:ethical-alignment-score}
    \textbf{Principle} & \textbf{Features} & \textbf{Definition} \\
    \midrule
    Temporary   & Clearly defined lifetime and documented decommissioning process; limited data retention & The scope of the application is limited to the current pandemic and no data is stored longer than deemed necessary to perform reliable and accurate contact tracing. \\
    
    Consensual & Opt-in with informed consent for download, use, sharing of data; not tied to other benefits & Download and use the application is entirely voluntary and not tied to benefits such as access to healthcare or other services; application practices informed consent for any significant events involving a user's data.   \\
    
    Minimal & Minimal data is collected; only one form of tracking at minimally granular level  & Only uses BLE or GPS for purposes of tracking contacts with no additional tracking; no personally identifiable information (PII) collected. \\
    
    Transparent & Open source code, documentation, and privacy policy with clear explanations & All aspects of the application should practice high levels of transparency in the form of code, documentation, and policies in clearly articulated language. \\
    
    Equal Access & Freely available to all with no paywall & Application should be freely available for Android and iOS (in countries where iPhones are common) with no associated costs or contingencies. \\
    
    Data Ownership & Decentralized storage; control of data lies with user; encrypted and secured data & An individual's data should not be kept in a centralized storage system and individual's should have full control of their data at all times; data should be secured and encrypted during transmission and at rest. \\
    
    Re-identification & Fully anonymous; uses decentralized matching (DP-3T, Apple/Google, TCN) & An individual's data is matched in a decentralized manner (on user devices) and is fully anonymous such that no individual can be re-identified if their data is compromised.\\
    
    Accuracy & Application uses accurate technology for contact tracing & Bluetooth/Bluetooth Low Energy (BLE) used for establishing contacts and only test-verified cases reported.\\
    \hline
    \vspace*{0.2in}
\end{tabulary}
\end{table}

To evaluate contact tracing applications based on the criteria provided in the previous section, we propose the concept of an "Ethical Alignment Score". The Ethical Alignment Score of an application is a value in the range 0-8 representing the extent to which the application meets principles of ethical contact tracing, as outlined in table \ref{table:ethical-alignment-score}. A score of zero indicates that the application fails to even partially fulfill any of the criteria, while a score of eight is the ideal case indicating that the app conforms with all principles of ethical contact tracing that we were able to assess under this framework. Each principle in table \ref{table:ethical-alignment-score} corresponds to a single point in the app's score. The full dataset, as well as an interactive map showing all the applications, is available here: \url{https://benjaminlevy.ca/covid-apps}. This dataset also further breaks down each principle into 2 or 3 more specific sub-principles, on the basis of which we evaluated each app. We hope to keep this dataset updated as new applications are developed and released, as well as when new information comes to light about how these apps and protocols function. As such, it is possible that the apps displayed in the interactive map and data table may be different from the ones used in the present analysis, since only those two elements will be updated over time. All analyses were completed using Python v3.8.2.
\hypertarget{results}{%
\section*{The state of the contact tracing ecosystem}\label{results}}

We collected data on a total of 152 applications from 77 different countries (counting the EU as a separate country for apps that are EU-wide). Of these, 124 are released on Apple's App Store or the Google Play Store (or are available as web-based platforms), while 17 have been cancelled and 11 are still in development at the time of writing. The app release dates span from mid-February (Singapore's TraceTogether was released on February 18, 2020) until January (Louisiana's COVID Defense was released on January 18, 2021).

\begin{figure}[h]
    \centering
    \includegraphics[width=\textwidth]{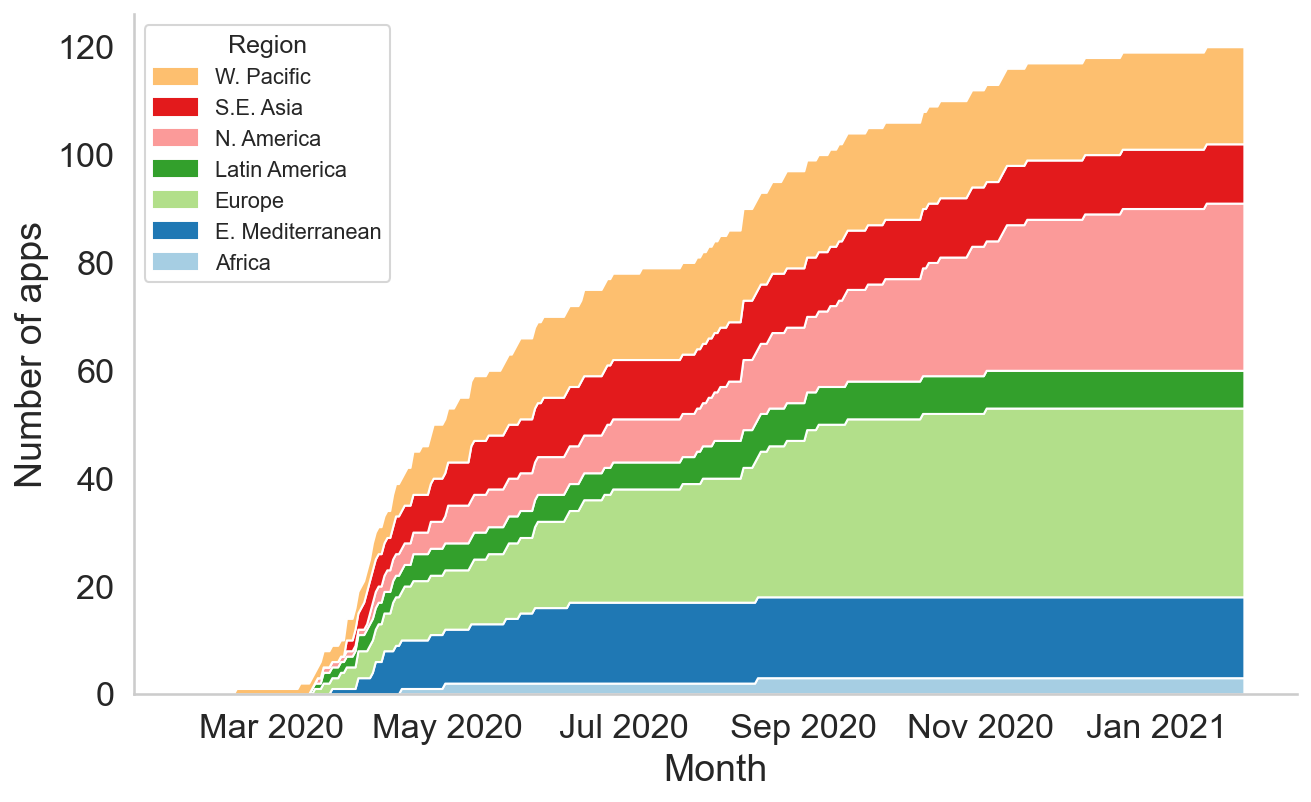}
    \caption{Released contact tracing apps, grouped by WHO region. North America (U.S. and Canada) is separated from Latin America, although normally the WHO groups both into a single Americas region. The total number of apps reflects only those apps that could be assigned to a single country or region.}
    \label{fig:apps-by-region}
\end{figure}

Figure \ref{fig:apps-by-region} shows the progression of app releases in each WHO region (with Canada and the US separated out from Latin America in the North America region). Initially, most new applications came from the Western Pacific region (including Singapore), Southeast Asia, and the EU. There has since been an explosion of apps released across the world, particularly in the EU, Eastern Mediterranean, and North America.

\hypertarget{results-protocols}{%
\subsection*{Comparison of contact tracing/exposure notification protocols
}\label{comparison-of-protocols}
}

\begin{figure}
    \centering
    \includegraphics[width=\textwidth]{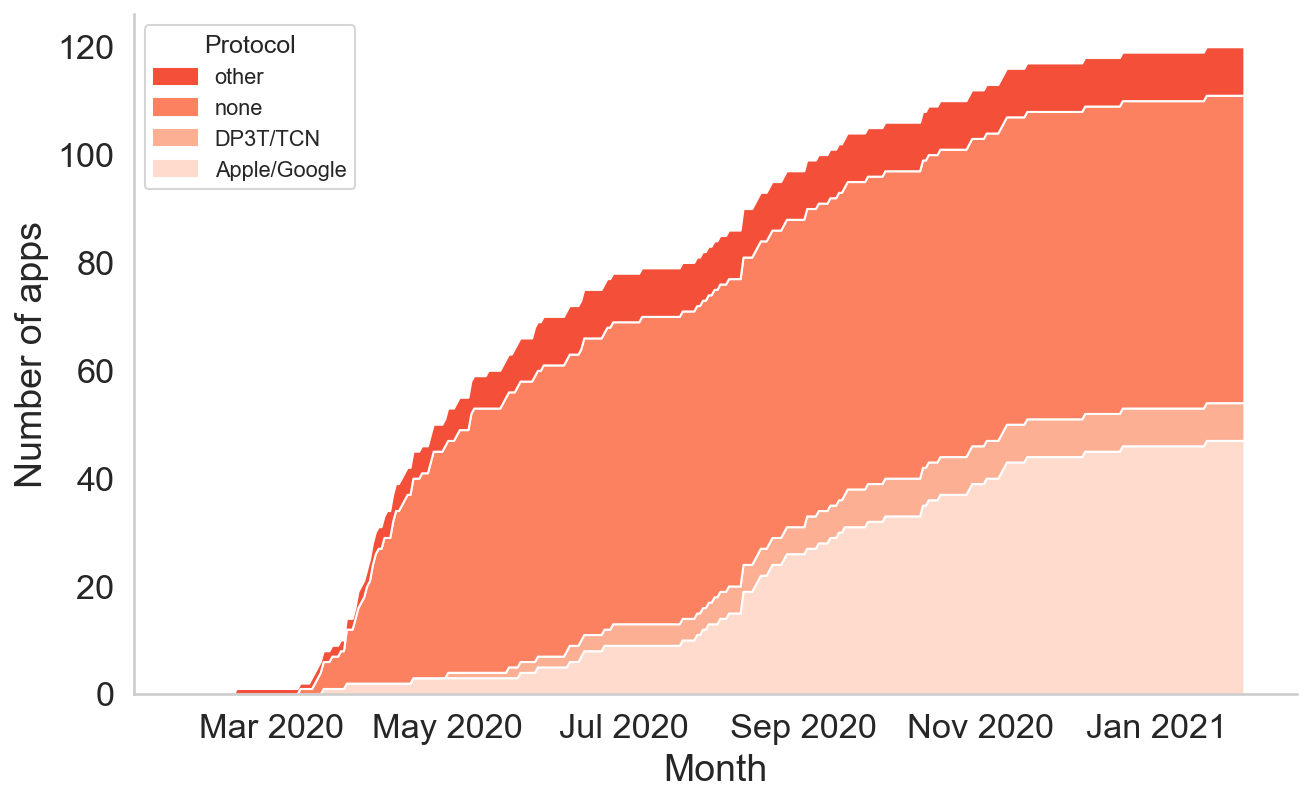}
    \caption{Number of contact tracing apps implementing different protocols. The major protocols include Apple/Google, and the closely-related DP3T/TCN, both of which are decentralized protocols. Other protocols are combined into a single category for brevity, and include OpenTrace, SafePaths, ROBERT, PEPP-PT, ReCoVer, Whisper, and Safe2.}
    \label{fig:apps-by-protocol}
\end{figure}

The choice of contact tracing protocol (or lackthereof) has varied substantially throughout the course of the pandemic. Early on, most apps did not employ an established protocol. By the beginning of September, this had shifted such that the majority of new applications employed the Apple/Google exposure notification protocol (see figure \ref{fig:apps-by-protocol}).

\begin{figure}[h]
    \centering
    \includegraphics[width=\textwidth]{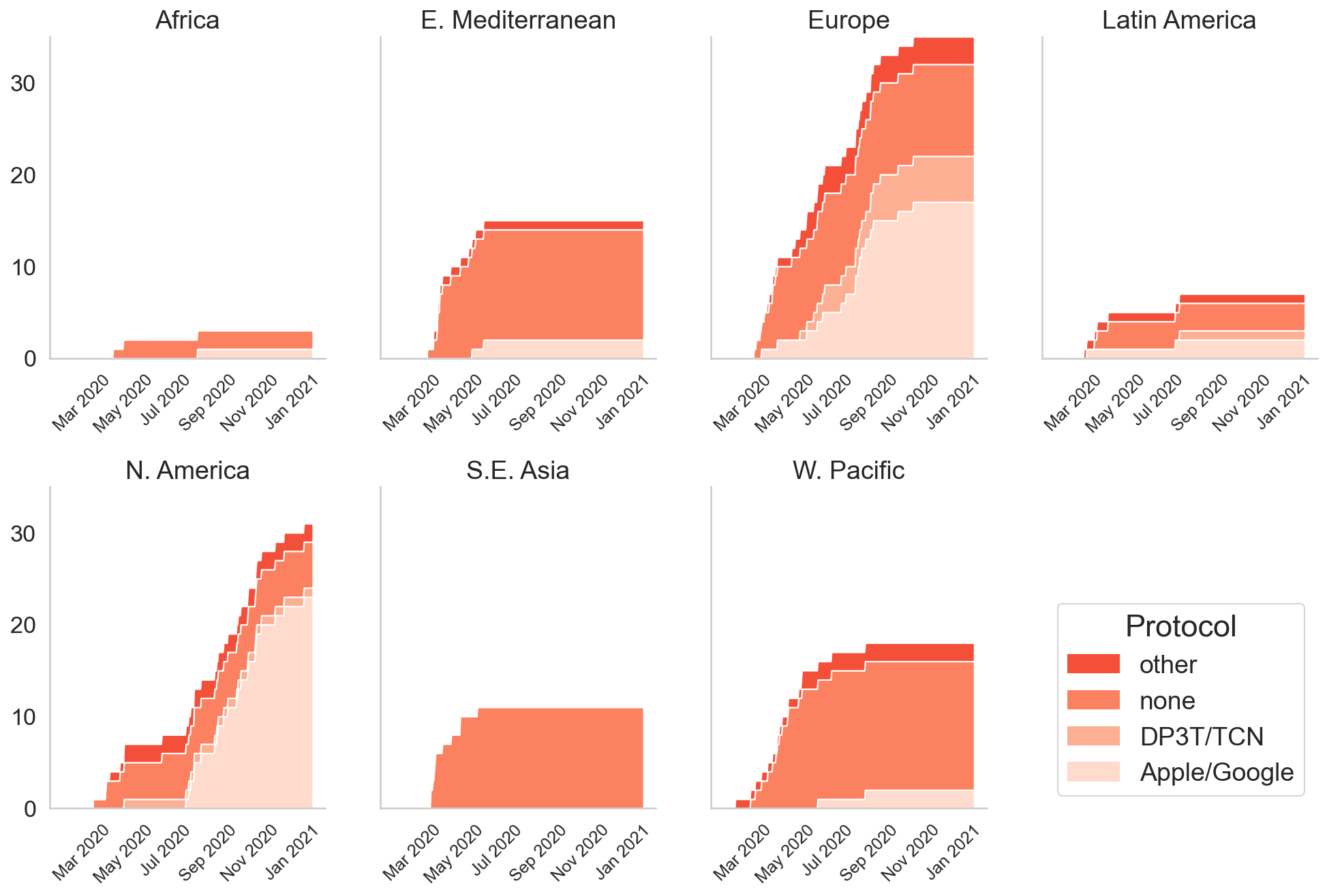}
    \caption{Number of released contact tracing apps implementing different protocols, disaggregated by WHO region.}
    \label{fig:apps-by-protocol-region}
\end{figure}

Figure \ref{fig:apps-by-protocol-region} disaggregates the apps by region and protocol. The dominance of the Apple/Google exposure notification protocol later in the pandemic is mainly localized to Europe and North America, whereas other regions, such as Southeast Asia, Western Pacific, and Eastern Mediterranean mainly developed apps without a protocol early in the pandemic.

\hypertarget{comparison-of-government-applications}{%
\subsection*{Comparison of Government-based Applications}\label{comparison-of-government-applications}}

\begin{figure}[h]
    \centering
    \includegraphics[width=\textwidth]{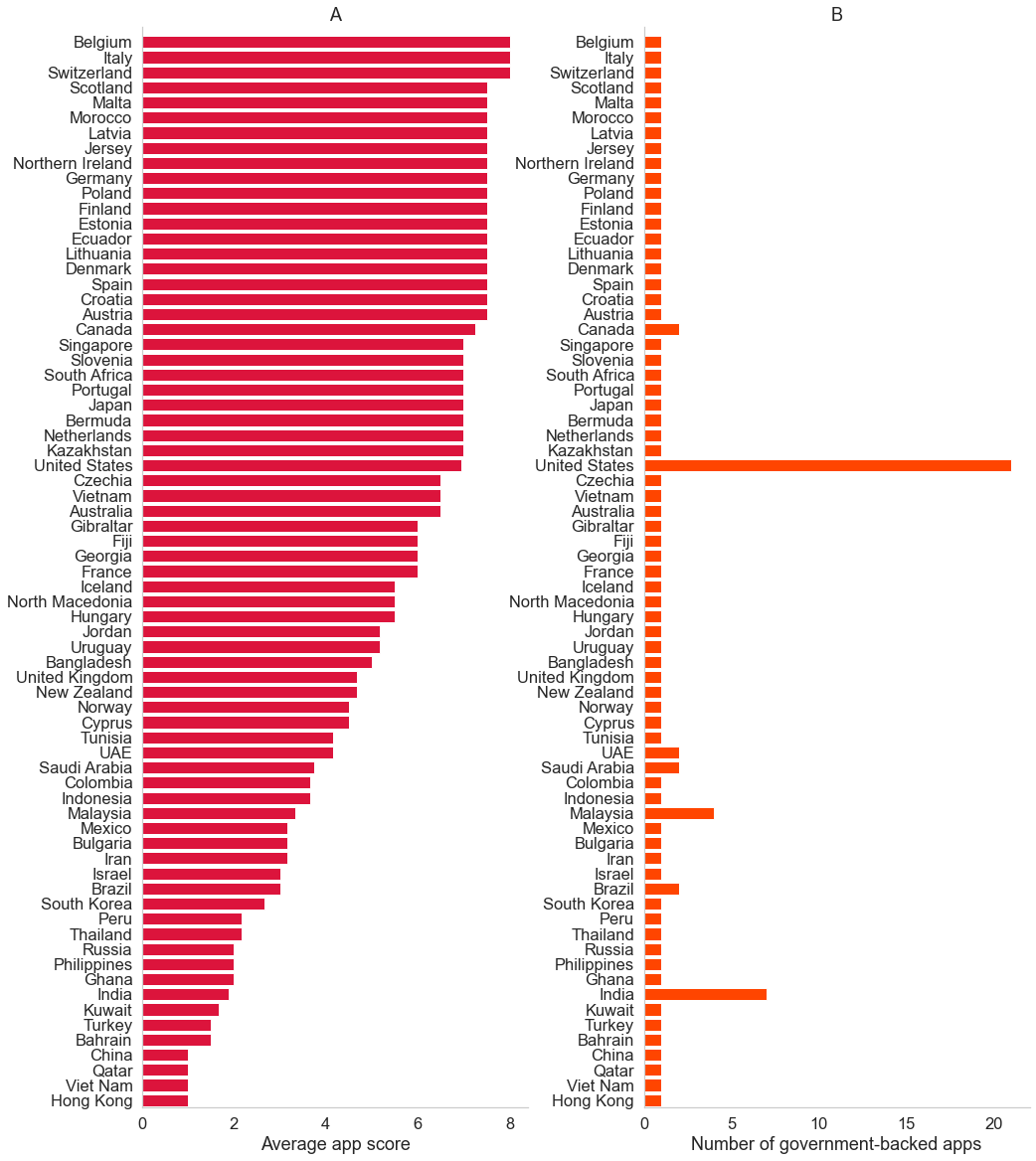}
    \caption{(A) Average Ethical Alignment Score for each country with a government-backed app. (B) Number of government-backed apps per country. Countries whose apps conform most closely with principles of ethical digital contact tracing include Switzerland, Italy, and Belgium. Countries/Territories with the lowest average score include Hong Kong, Viet Nam, Qatar, and China. The United States and India far outpace other countries in terms of how many government-backed apps have been released.}
    \label{fig:government-apps}
\end{figure}

110 applications (72\%) were developed or backed by government entities. The corresponding ethical alignment scores for these applications, averaged by country, are provided in ranked order in Figure \ref{fig:government-apps}. Switzerland, Italy, and Belgium were among the countries with the highest average scores (8), while Hong Kong, Vietnam, Qatar, and China received the lowest scores (1). Although most countries have a single government-backed application, two notable outliers are the United States and India, with 24 and 7 government-backed apps released, respectively. Whereas the United States' apps were released late in the pandemic (all on or after August 3\textsuperscript{rd}, 2020) all of India's apps were developed relatively early, within the period from March 7\textsuperscript{th} to May 9\textsuperscript{th}.

Common factors that resulted in lower alignment scores included (1) the use of a centralized database to store user data, (2) GPS tracking of individuals, and (3) lack of informed consent associated with the collection, storage, or use of individual data. While none of these applications received the worst possible score of zero, only 3 applications received a perfect score of 8. Among the applications with near-perfect scores (7.5), the most common reason for losing points was not having a clearly defined lifetime for the application (13/15).

\hypertarget{comparison-of-non-government-applications}{%
\subsection*{Comparison of Non-Government-Based Applications}
}

In our dataset, 42 applications were not directly associated with a governmental entity. Of these, 20 have been released, 15 have been cancelled, and 7 are either still in development or have an indeterminate status. The countries with the most non-governmental applications of any operational status are the United States (11), Germany (8), Canada (4), the Philippines (2), and Italy (2). Most non-governmental applications were released before June (62\%) and all were released before July. When compared to the government-backed applications, these non-governmental applications score lower on average in almost every question (see table \ref{tab:gov-vs-nongov}), with the exception that slightly more non-governmental apps are open source (30\%) than governmental apps (28\%).

\begin{table}[h]
    \centering
    \caption{Comparison of government-backed vs. non-governmental applications; only released apps. Figures are reported as total count (\%) for individual questions and average for the score.}
    \begin{tabular}{lcc}
        \toprule
        {} & Non-government &  Government \\
        Question                                      &                &             \\
        \midrule

            1.1. App has defined lifetime                 &         0 (0\%) &    17 (16\%) \\
            1.2. Data storage is time-limited             &       10 (50\%) &    75 (72\%) \\
            2.1. Opt-in download and use                  &       14 (70\%) &    88 (85\%) \\
            2.2. Opt-in data sharing                      &       14 (70\%) &    86 (83\%) \\
            2.3. Not tied to other benefits               &        7 (35\%) &    68 (65\%) \\
            3.1. Data used only for establishing contacts &        7 (35\%) &    65 (62\%) \\
            3.2. No PII collection                        &        4 (20\%) &    57 (55\%) \\
            4.1. Open source                              &        6 (30\%) &    29 (28\%) \\
            4.2. Published privacy policy                 &       13 (65\%) &    80 (77\%) \\
            5.1. Freely available                         &       16 (80\%) &  104 (100\%) \\
            5.2. Android and iOS (when necessary)         &       16 (80\%) &  104 (100\%) \\
            6.1. Decentralized storage                    &        7 (35\%) &    69 (66\%) \\
            6.2. Can erase data                           &       11 (55\%) &    71 (68\%) \\
            7.1. Decentralized matching                   &        4 (20\%) &    55 (53\%) \\
            7.2. Rotating randomized beacon               &        5 (25\%) &    64 (62\%) \\
            8.1. Contact accuracy (BLE, not GPS or other) &        6 (30\%) &    69 (66\%) \\
            8.2. Positive cases verified by test          &        6 (30\%) &    76 (73\%) \\
            {\bf Average score}                           &      {\bf 4.31} &   {\bf 5.31} \\

        \bottomrule
    \end{tabular}
    \label{tab:gov-vs-nongov}
\end{table}

\hypertarget{comparison-of-time-evolution}{%
\subsection*{Time Evolution of Ethical Alignment}\label{comparison-of-protocols}}

\begin{figure}[h]
    \centering
    \includegraphics[width=\textwidth]{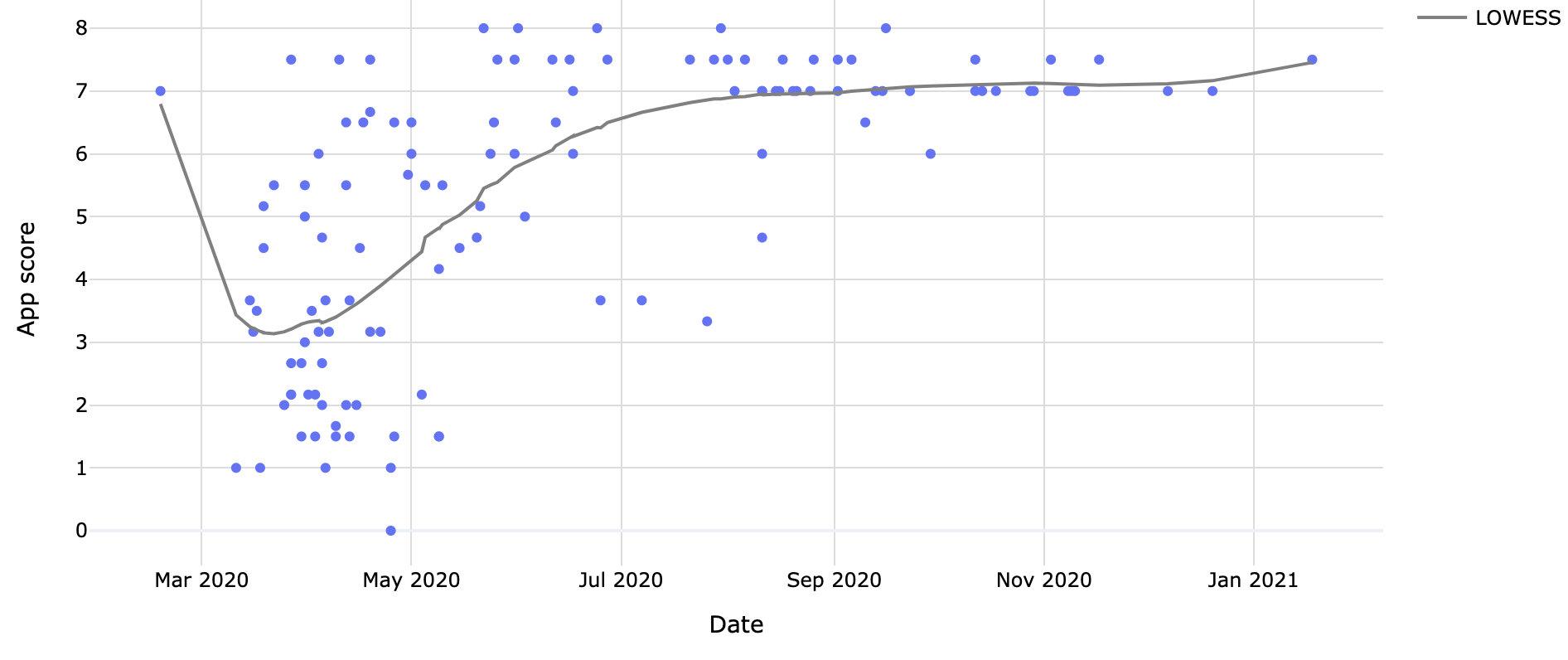}
    \caption{Ethical alignment scores of applications as a function of release date. The grey line indicates a LOWESS curve fitted to the data.}
    \label{fig:time-evolution}
\end{figure}

To distinguish any trends in ethical alignment scores over time, we examined the ethical alignment scores of different applications as a function of application release dates. As shown by Figure \ref{fig:time-evolution}, there was initially high variation in scores among apps, while later in the pandemic, apps have tended to have much higher ethical alignment scores.

\begin{figure}[h]
    \centering
    \includegraphics[width=\textwidth]{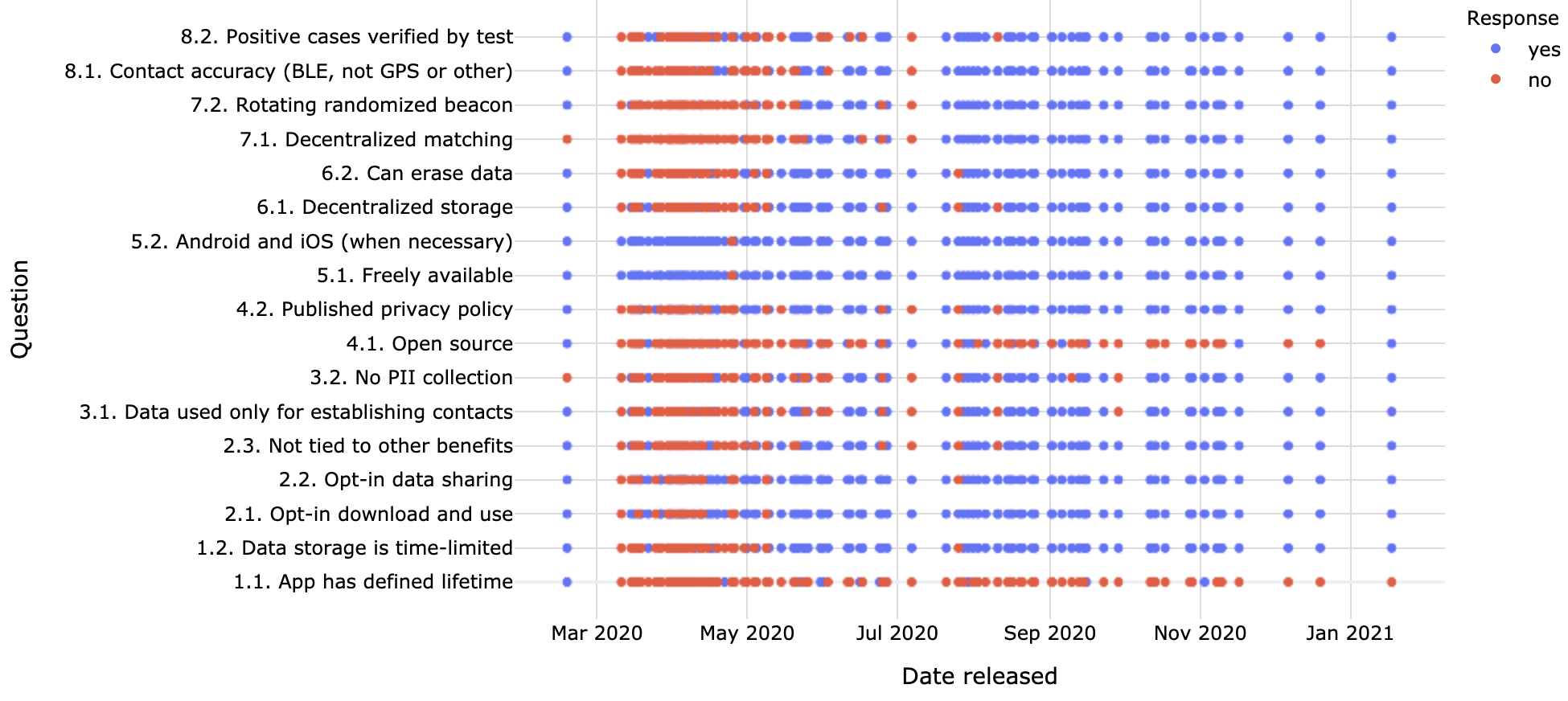}
    \caption{App scores disaggregated by question and release date. Each column represents a single application, while each dot represents the value of that application for each question (rows); blue indicates the app satisfies the criterion, red indicates the app does not satisfy the criterion.}
    \label{fig:apps-by-question}
\end{figure}

Disaggregated scores can be seen in Figure \ref{fig:apps-by-question}. Two criteria that are consistently less satisfied than the others regardless of the date are whether the app is open source and whether the app has a clearly defined lifetime.

\hypertarget{us-apps}{
\subsection*{Apps in the United States}\label{us-apps}}

\begin{figure}[h]
    \centering
    \includegraphics[width=\textwidth]{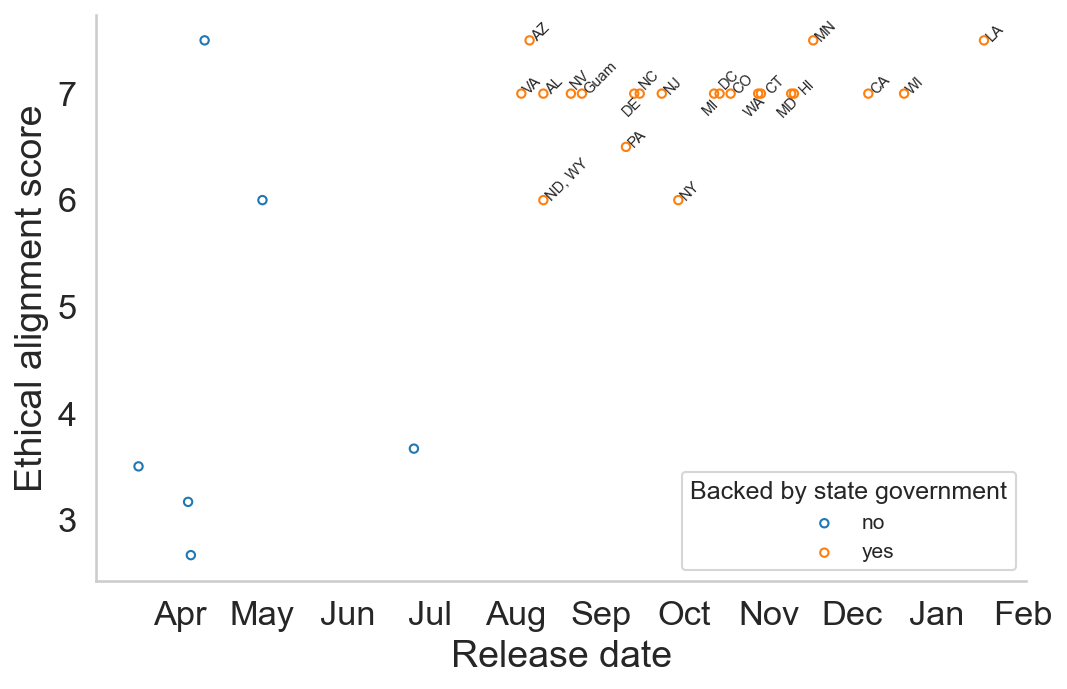}
    \caption{Ethical alignment scores vs. release date for contact tracing apps released in the United States. Color denotes government (orange) or non-government (blue).}
    \label{fig:us-apps}
\end{figure}

Apps in the United States can be neatly divided into two groups shown in Figure \ref{fig:us-apps}: (1) private apps released early in the pandemic and (2) government-backed apps released late in the pandemic. The large number of government-backed apps in the US relative to other countries is due to the fact that apps have been developed at the state level, as opposed to the national level in most other countries.

\hypertarget{analysis-limitations}{%
\subsection*{Limitations of Analysis}\label{comparison-of-protocols}}

The assessment of contact tracing applications was limited in scope by the public availability of data which made analysis of some action points from the WHO ethical contact tracing guidelines difficult. In an effort to counter this, emails were sent to application developers for all applications examined in this study, where possible, with a 17-point questionnaire covering the items that were not able to be assessed using public information. The response rate to this questionnaire was approximately 20\%, which was deemed too low to make generalizations due to the possibility of self-selected sampling.

Because of the international scope of this work, many apps did not have English-language documentation readily available. Although we attempted to mitigate this with Google Translate, it is likely that certain aspects of these apps were lost in translation.

Finally, it may be the case that apps that implement the Apple/Google exposure notification protocol are overrepresented in the dataset relative to non-Apple/Google apps. This is likely due to the presence of lists of apps that implement the Apple/Google protocol, such as \href{https://developers.google.com/android/exposure-notifications/apps}{one curated by Google itself}. Conversely, apps that do not implement the protocol may be harder to find, which would contribute to their apparent scarcity later in the pandemic.
\hypertarget{discussion}{%
\section*{Discussion and significance}\label{discussion}}

In this paper, we have provided a synthesis of existing databases of contact tracing applications, supplemented with additional apps found in the course of research, creating, to our knowledge, the most comprehensive collection and evaluation of COVID-19 contact tracing applications to date. Figure \ref{fig:apps-by-region} indicates the existence of several waves of contact tracing applications. The first applications were developed in March 2020, coinciding with the first wave of national lockdowns around the world. The period of April-May 2020 saw a precipitous rise in applications, somewhat evenly split between WHO regions, with the exception of the United States and Canada. The period of June saw the release of several more applications, but this quickly stagnated in July until an influx of new applications, predominantly from North America and Europe, during the period August-November. The cause of the temporary stagnation during July is unclear. However, it may be related to incorporating privacy-preserving contact tracing into new applications following the release of the Apple-Google implementation in late May 2020, as well as the possibility that our data collection procedure may have missed some apps that were released in this period.

Figure \ref{fig:government-apps} illustrates the wide variance in country-level differences in ethical alignment scores for government-backed applications. We advance several possible explanations for these disparities. First, there may be intrinsic reasons why a country's apps scored higher, such as higher cultural value placed on privacy vs. public health. Similarly, the large number of high-scoring apps implementing the Apple/Google protocol in European and North American countries may indicate greater cultural appropriateness of that particular protocol to Western nations or greater appeal of the protocol to Western governments. Second, the variance in ethical alignment between countries may also reflect a government's speed of innovation. Figures \ref{fig:time-evolution} and \ref{fig:apps-by-question} show an apparent disparity between the ethical alignment scores of early implementations, released before June 2020, and those released subsequently. The disparity bisects applications into two opposing categories: data-centric and privacy-centric \citep{fahey2020covid}. Early implementations tended to exhibit lower ethical alignment scores and placed more emphasis on data collection than individual privacy. In contrast, later implementations tended to exhibit higher ethical alignment scores, emphasizing individual privacy as well as greater transparency. This transition roughly coincided with the Apple-Google contact tracing implementation in April 2020 \citep{appleGoogleContactTracing}, which upon release was touted by some as well-formulated privacy-preserving contact tracing \citep{acluAppleGoogle}. By September 2020, applications reached an asymptote corresponding to an ethical alignment score of 7, on average. Therefore, paradoxically, the governments that were {\it more} innovative and willing to experiment with digital technologies for combating COVID-19 may have been more likely to implement apps with sub-optimal privacy features, whereas governments that waited to roll out their own contact tracing application may have benefited from the development of privacy-preserving protocols like the Apple/Google Exposure Notification protocol.

Both Germany and the United States are notable for having a large number of non-governmental contact tracing applications relative to other countries, with 8 and 11, respectively. In Germany, many of these applications were developed as part of the WirVsVirus hackathon, which took place from March 20\textsuperscript{th} to 22\textsuperscript{nd} and whose goal was to bring together coders to rapidly prototype digital tools to help confront the challenges posed by COVID-19, including contact tracing applications \citep{WirVsVirus}. In the United States, the significant number of independent applications may be a result of both a huge capacity to produce digital solutions (e.g. Apple and Google, the producers of the eventually dominant Exposure Notification protocol, are located in the United States) and a hesitancy of governmental authorities to develop or sanction their own apps. Figure \ref{fig:us-apps}, which contrasts different contact tracing implementations in the United States, shows a clear distinction between government-backed and non-governmental contact tracing implementations. Non-governmental applications tended to exhibit lower ethical alignment scores (as was the case across the world; see table \ref{tab:gov-vs-nongov}) and were developed significantly faster than government-backed applications, with a three-month gap between the release of the last non-governmental application and the first government-backed application. This delayed response may be due to a general mistrust of public health measures or technologies that have the capacity to infringe privacy or track individuals. According to the Best Countries survey conducted by U.S. News, the United States ranks near the bottom in terms of the citizenry's trust in the government with respect to matters of healthcare \citep{usNewsTrust}. Another possible explanation is the generally decentralized nature of the United States' public health infrastructure, illustrated by the fact that multiple apps have been rolled out independently by different states, as opposed to a single federal app.

\hypertarget{future-implications}{%
\subsection*{Future Implications}\label{future-implications}}

The evolution of these applications throughout the COVID-19 pandemic has several important implications on the future of large-scale epidemiological monitoring. If effective, such a large-scale system could prevent or contain large numbers of infections in the event of an outbreak while avoiding the worst impacts of an indiscriminate lockdown. Conversely, the same system could be used under normal circumstances to detect and monitor yearly flu outbreaks. Monitoring flu outbreaks would invariably save lives, but the frequent mutation of flu strains also provides a convenient justification to perpetuate the existence of contact tracing applications without end. In this scenario, the potential exists for governments and companies to abuse contact tracing to perform mass surveillance of civilians using seemingly justified means. The implementation of ethical contact tracing minimizes this possibility by empowering civilians with control over the collection and disclosure of their data and ensuring radical transparency, accountability, and data security. All of these factors are instrumental in building public trust, which, if the situation ever arises, would be minimally necessary for an environment where some degree of digital surveillance is required for public protection against the outbreak of an infectious disease.

Another implication of these analyses is the way in which standards around privacy and ethics are set. In lieu of a central authority to enforce strong standards early in the pandemic, Apple and Google effectively became the arbiter of ethical contact tracing, in some instances indirectly pressuring governments to adopt their protocols (in the case of the UK's NHS app) \citep{sabbagh_hern_2020} or outright banning apps that did not conform to their standards (in the case of Iran's ac19r app) \citep{mitTechRev2020}. Although Apple and Google's protocol did enforce strong protections for individual rights, in our view, it is crucial that future decisions around public health technologies be made in a way that involves multiple stakeholders, including public health authorities, elected representatives, and the general public.
\hypertarget{conclusion}{%
\section*{Conclusion}\label{conclusion}}

Now that vaccines have been approved for widespread distribution, we are likely to see a gradual return to the normalcy of public life and an easing of government restrictions. However, the virus will likely remain with us still for an indefinite period until a sufficient level of immunity is reached through vaccination. Using mobile phone data to track, trace, and quickly respond to new infections and flare-ups has drawn much attention as a potential solution. However, high levels of public adoption are critical to the success of contact tracing applications. Since coercing the public into using these applications would constitute an unacceptable violation of several critical ethical principles, the main avenue to promote adoption is building public trust. It is vital that the public feel that they are in control of their data at all times. Thus, we are heartened to see that most applications -- especially those released later in the pandemic -- use non-coercive, opt-in, decentralized, and anonymous contact tracing schemes.

It is arguably still too early to tell whether the use of the contact tracing applications discussed here has altered the course of the pandemic in a meaningful way, especially since many applications have been released somewhat recently. However, it is clear from the data presented here that there has been significant heterogeneity, both within and between regions, and across the evolution of the pandemic, in terms of the ethicality of these applications. Crucial to investigating whether these apps have proved efficacious or not would be the release of precise download numbers from the app developers, which would allow researchers to determine whether uptake reached the thresholds theorized to ameliorate the spread of infection. We hope that future work can build off the database we have assembled to further probe whether these apps have been helpful and how the use of similar technology in future public health crises can be better managed.
\hypertarget{conclusion}{%
\section*{Acknowledgments}\label{Acknowledgments}}

We would like to thank Professors Jim Waldo and Mike Smith of the School of Engineering and Applied Sciences at Harvard University for their guidance during the initial stages of this research work, as well as Andrés Colubri of MIT, for his assistance in obtaining relevant data on contact tracing applications.

\subsection*{Disclosure Statement}
The authors have no conflicts of interest to declare.

\printbibliography

\end{document}